\documentclass[
    ,final            
  ]
  {aipproc}
\usepackage{graphicx}
\layoutstyle{8x11single}
\pdfoutput=1
\begin{document}

\title{The evolution of massive binaries in a dense stellar cluster environment}

\classification{97.10.Cv, 97.10.Gz, 97.10.Kc, 97.10.Me}
\keywords      {Binary evolution, mass loss by stellar winds, rotation, cluster stellar dynamics}

\author{Dany Vanbeveren}{
  address={Astrophysical Institue, Vrije Universiteit Brussel, Pleinlaan 2, 1050 Brussels, Belgium}
}

\begin{abstract}

We first consider binary processes and focus on the effect of rotation on binary evolution and on the mass transfer during Roche lobe overflow. The second part highlights the importance of close pairs on the comprehension of the evolution of stellar populations in starburst regions.

\end{abstract}

\maketitle

\section{Introduction}

 Massive stars are among the most important objects in the Universe and many (most?) of them are formed in binaries.  A selection of observational and theoretical facts that illustrate the importance of binaries and the evolution of massive and very massive stars in clusters with special emphasis on massive binaries have been summarized in two recent review papers  (Vanbeveren, 2009,  2010).  The present paper can be considered as an addendum of both reviews.

\section{The evolution of massive binaries}

The main differences between single star evolution and the evolution of the same star when it is a binary component are related to  the Roche lobe overflow (RLOF) process and to binary processes which determine the rotation rate of the star. 
 
\subsection{Rotation and binaries}

Due to tidal interaction the massive primaries of most of the binaries are expected to be slow rotators. Only in very short period systems (P = 1-2 days) it can be expected that massive primaries are rapid rotators for which the evolution proceeds quasi-homogeneously (De Mink, 2010). 

In binaries where the RLOF of the primary is accompanied by mass transfer towards and mass accretion onto the secondary, the secondary spins-up and very rapidly rotational velocities are reached close to the critical one (Packet, 1981). This happens in  systems where the RLOF occurs when the outer layers of the primary are mainly in radiative equilibrium (Case A / Case Br systems). Population synthesis predicts that many Be stars are formed this way (Pols and Marinus, 1994; Van Bever and Vanbeveren, 1998). The latter two studies illustrate that one may expect many Be stars with a subdwarf (sdO) or white dwarf (WD) companion. The high temperature of these companions makes them very hard to detect and this may be the reason why so few are known at present ($\phi$ Per is an exception). The Be-components in Be-X-ray binaries form an interesting subclass of the Be sample because here we have all reasons to believe that binary action has been important in the formation of the Be star. Many Be single stars are also expected to form via binary mass transfer. The reason is that the supernova explosion of a massive primary disrupts the binary in most of the cases. This means that the fact that many Be stars have a neutron star companion means that even more Be single stars have had a similar past as the Be stars in the Be-X-ray binaries.

The optical components of the standard high mass X-ray binaries are former binary secondaries where mass and angular momentum accretion may have occured. The mass and helium discrepancy for single stars discussed by Herrero et al. (1992) is also visible in the optical component of the X-ray binary Vela X-1 (Vanbeveren et al., 1993) and we proposed  {\it the accretion induced full mixing model}.  The idea was the following: due to mass and angular momentum accretion, mass gainers spin-up. This may induce efficient mixing. We simulated this possibility with our evolutionary code by fully mixing the mass gainer and, after the mass transfer phase, following the further evolution of the mixed star in a normal way. In this way we were able to explain the helium and mass discrepancy in Vela X-1. The more sophisticated mass gainer models of Cantiello et al. (2007) demonstrate that our simplified models are not too bad.

The rotational velocity distribution of O-type stars (Vanbeveren et al., 1998a) illustrates that many O-type stars are relatively slow rotators, corresponding to an initial average rotational velocity for O-type stars of $\le$ 200 km/s for which indeed the effect of rotation on their evolution is rather modest (Maeder and Meynet, 2000a). The distribution also shows that there is a subset of very rapid rotators. However, many of these rapid rotators are runaway stars (they have a space velocity $\ge$ 30 km/s) and this may indicate that these stars were former binary components, e.g. they became rapid rotators due to the mass transfer process in a binary and they became runaways due to the supernova explosion of their companion, or they became rapidly rotating runaways due to stellar dynamics in dense stellar clusters, in which case they were former binary members as well but their formation was governed by star merging. An interesting test bed for this type of process may be $\zeta$ Pup which is indeed a rapidly rotating runaway. Note that Mokiem et al. (2006) obtained rotational velocities of 21 OB dwarfs in the SMC and concluded that the average  $v_{rot}$=160-190 km/s. Since massive dwarfs are stars close to the zero age main sequence, this average value is indicative for the average initial rotation velocity of OB-type stars. Remark that this value is not significantly different than the initial average value of Galactic O-type stars whereas, similar as in the Galactic sample, the most rapid rotators in the SMC are runaway stars. 

All in all, it looks to me that rotation is important for massive star evolution but perhaps mainly within the framework of binaries in combination with stellar dynamics in young dense clusters.  

\subsection{The Roche lobe overflow process}

When the RLOF starts when the mass loser has a convective envelope (Case Bc and Case C), the mass loss process happens on the dynamical timescale and a common envelope forms. It can be expected that the common envelope is lost as a superwind where most  of the energy is supplied by orbital decay and it stops when the two components merge or, when merging of the two stars can be avoided, when most (but not all) of the hydrogen rich layers of the mass loser are removed. This phase is so rapid that it is unlikely that  mass accretion plays an important role for the evolution of the secondary star and therefore the latter may not become a rapid rotator.  

When the RLOF starts when the mass loser has a radiative envelope (Case A and Case Br), the mass loss process happens on the Kelvin-Helmhotz time scale of the loser and when the initial mass of the gainer is not too much smaller than the initial mass of the loser, mass transfer and mass accretion becomes possible. 

The evolution of the mass loser in Case Br and in most of the Case A massive binaries is very straightforward: due to RLOF the redward evolution of the loser is avoided (e.g., massive primaries in Case A or Case Br binaries do not become red supergiants). The RLOF stops when the loser has lost most (but not all) of its hydrogen rich layers and helium starts burning in the core. At that moment the loser resembles a WR-like star (when the mass is large enough the WR-like star is expected to be a guinine WR star with hydrogen in its atmosphere, typically X = 0.2-0.3). 

The evolution of the mass gainer in Case A and Case Br binaries is governed by mass and angular momentum accretion and rotation plays a very important role (see the previous subsection).

An important question is whether or not the RLOF in Case A or Case Br binaries is quasi-conservative. Let me first remark that removing matter out of a binary at a rate which is similar to the rate at which the primary loses mass, requires a lot of energy, much more than the intrinsic radiation energy of the two components which is in most cases only sufficient in order to drive a modest stellar wind. The Utrecht group promoted a massive binary model where extensive mass loss from the system happens when the mass gainer has been spun-up by mass and angular momentum transfer and reaches a rotational velocity close to the critical one (Petrovic et al., 2005; De Mink, 2010). Both studies use a first attempt to link rotation and mass loss proposed by Langer (1997) but Glatzel (1998) showed that the proposed relation may be not correct due to the fact that it does not account for the effect of gravity darkening (von Zeipel, 1924). An alternative and attractive formalism has been derived by Maeder and Meynet (2000b) where the effect of gravity darkening was taken into account. This relation demonstrate that for most of the massive stars (with an average initial rotational velocity of $\approx$ 200-300 km/s) the increase of the stellar wind mass loss with respect to the non-rotating case is very modest. The increase is significant for stars with a large Eddington factor $\Gamma$ (e.g., stars with an initial mass $\ge$ 40 M$_\odot$) that are rotating close to critical. However, the following remarks are appropriate: there are no observations yet to sustain the relation proposed by Maeder and Meynet (Puls et al., 2010). Even more, one may wonder whether or not rotation can be a significant mass loss driver, since even at critical rotation, the rotational energy is at most half the escape energy of a massive star (Owocki, 2010, private communication). Van Rensbergen et al. (2008) proposed the following model: the gasstream during RLOF forms a hot spot either on the surface of the star when the gasstream hits the mass gainer directly, or on the Keplerian disc when mass transfer proceeds via a disc. The radiation energy of the hot spot in combination with the rotational energy of the spun-up mass gainer can then drive mass out of the binary. The following illustrates that this model may not work. For the sake of simplicity, let us neglect rotation because as was stated already before, rotation is not an efficient mass loss driver whereas even at critical break up, the rotation energy is too small compared to the escape energy of a massive star. The radiation energy $L_{acc}$ generated by the accretion of the gasstream is given by

\begin{equation}
L_{acc} = G\frac{\dot{M}_{acc}\,M}{R}
\end{equation}

\noindent where we neglect the fact that the gasstream does not originate at infinity but at the first Lagrangian point (it can readily be checked that this assumption does not significantly alter our main conclusions). $\dot{M}_{acc}$ is the mass accretion rate, M and R are the mass and the radius of the gainer. When $\eta$ is the fraction of $L_{acc}$ that is effectively transformed into escape energy, it follows that

\begin{equation}
\eta\ L_{acc} = \frac{1}{2}\dot{M}_{out} v^{2}_{esc}
\end{equation}

\noindent with $\dot{M}_{out}$ the binary mass loss rate and $v_{esc}$ the escape velocity. When we only account for the escape energy of the mass loser, the foregoing equations result into

\begin{equation}
\dot{M}_{out}=\eta\ \dot{M}_{acc}
\end{equation}

\noindent Detailed hydrodynamic calculations of stellar winds reveal that the efficiency factor for converting radiation energy into kinetic energy is of the order of 0.01 and 0.001 (Nugis \& Lamers, 2002). So, unless the efficiency is much higher, equation (3) then illustrates that in general the accretion energy may cause mass loss out of a binary but this loss is much smaller than the mass accretion rate. Of course when the gainer rotates at the critical velocity, mass accretion on the equator will not happen. Possibly matter will pile up arround the gainer, accretion may happen on the rest of the star, or matter may leave the binary through the second Lagrangian point L2. When this happens the variation of the orbital period can be calculated in a straightforward way (e.g., Vanbeveren et al., 1998b). Note that in most of the population studies performed by different groups the effect of a non-conservative RLOF is investigated using this L2 model. Let me finally remark that mass that leaves the binary through the decretion/accretion disk of the gainer when the gainer rotates at the break up velocity, takes with the specific orbital angular momentum of the gainer but also the specific rotational angular momentum of the equator of the gainer. Interestingly, the sum of both momenta is roughly equal to the specific angular momentum of the L2 point.

\subsection{Some interesting observed test beds of the Roche lobe overflow process}

As discussed in the previous subsection, from theoretical considerations it is unclear whether or not Case A / Case Br evolution in massive binaries is quasi-conservative or not. Are there observed binaries that can be considered as RLOF test beds and allow us to say something about the RLOF? The best candidates are post-RLOF binaries or binaries at the end of RLOF where one can try to fit evolutionary models of binaries where the evolution of both components is followed simultaneously adopting different efficiency values for the mass transfer process. We did this for a number of Galactic massive binaries (Vanbeveren et al., 1998b) and de Mink et al. (2007) for SMC binaries. Here we reconsider an interesting system.

\subsubsection{RY Scuti}

The massive binary RY Scuti may be a key system for the discussion whether or not the RLOF in massive binaries is conservative. The spectroscopic study of Grundstrom et al. (2007) reveals that it is a O9.7Ibpe + B0.5I binary with a period = 11.2 days and masses 7 M$_\odot$ + 30 M$_\odot$. The O-type supergiant is the most luminous component which means that it is most probably a core helium burning star near the end of RLOF with a significantly reduced surface hydrogen content (X = 0.3-0.4). Similarly as the supergiant in HD 163181 (section 2.2.1) evolutionary calculations predict that the O-type supergiant will soon become a WR star. If the masses are correct then this system is an illustration of a massive binary where the RLOF was quasi-conservative for the following reasons. Evolution predicts that the 7 M$_\odot$ star comes from a star with initial mass $\le$ 20 M$_\odot$ that lost $\le$ 13 M$_\odot$ by RLOF. Since the initial mass of the secondary (mass gainer) must have been $\le$ 20 M$_\odot$ as well obviously, it must have accreted at least 10 M$_\odot$ of the $\le$ 13 M$_\odot$ lost by the loser in order to become a 30 M$_\odot$ star, e.g. the mass accretion efficiency must have been at least 80\% and we call this quasi-conservative. Note that the observations of Grundstrom et al.(2007) seem to indicate that there is some circum-binary material, but it is clear that the model discussed above does not contradict this fact.    

\section{Close pairs - key to comprehension of the evolution of stellar populations in starburst regions}

Mass transfer and mass accretion during a canonical RLOF in Case A/Br binaries  is responsible for the formation of a blue straggler sequence in young clusters (Pols and Marinus, 1994; Van Bever and Vanbeveren, 1998). Since these blue stragglers are mass gainers or binary mergers, it can be expected that they are rapid rotators, e.g., a cluster of slow rotators but with a significant population of close binaries will become populated with rapid rotators.     

Starburst99 is an interesting spectral synthesis tool to estimate all kinds of properties of starburst regions where only integrated spectra are available. However, it should be noted that this tool only accounts for the properties of single stars. The effects of binaries on the spectral synthesis of starbursts has been studied in Brussels: the effects of binaries on the evolution of W(H$_\beta$) (Van Bever \& Vanbeveren, 1999), the effect of binaries on the evolution of UV spectral features in massive starbursts (Belkus et al., 2003), the effect of binaries on WR spectral features in massive starbursts (Van Bever \& Vanbeveren, 2003), hard X-rays emitted by starbursts with binaries (Van Bever \& Vanbeveren, 2000). Note that Brinchmann et al. (2008) investigated the spectral properties of WR galaxies in the Sloan Digital Sky Survey and concluded that a comparison with theoretical population synthesis leeds to the conclusion that binaries are necessary.

\subsection{Intermezzo 1}

Vanbeveren (1982) discussed a possible relation between the maximum stellar mass in a cluster and the total cluster mass. It was concluded that {\it the integrated galaxial stellar IMF should be steeper than the stellar IMF}. This has been worked out in more detail about 20 years later by Kroupa and Weidner (2003) and Weidner and Kroupa (2006) who essentially arrived at the same conclusion.

A consequence of the fact that the mass of the most massive star in a cluster correlates with the cluster mass, is that it is possible that also the mass ratio distribution of the most massive binary population in the cluster correlates with the cluster mass. To illustrate, suppose that the cluster mass indicates that the maximum stellar mass/stellar object is 50 M$_\odot$, then one may expect binaries like 40 M$_\odot$ + 10 M$_\odot$ or 30 M$_\odot$ + 20 M$_\odot$ etc., but not 40 M$_\odot$ + 30 M$_\odot$.   
 
\subsection{Intermezzo 2}

Could it be that stars form in isolation? The origin of massive O-type field stars has been studied by De Wit et al., 2004. The authors proposed the following procedure in order to find a candidate O-type star that may have formed in isolation : take a non-runaway O-type field star and look for young clusters within 65 pc from the O-star. The value 65 pc was obtained by assuming that the lifetime of an O-type star is $\le$ $10^7$ yr. This is true for single stars however, the lifetime of an O-type star in a binary may be 2-3 times larger and therefore it may be necessary to look for young clusters within 200 pc. The model goes as follows: a 12 M$_\odot$ + 9 M$_\odot$ binary is dynamically ejected from a dense cluster with a velocity of 6 km/s. After 30 million years, when the binary is 200 pc away from its parent cluster, the 12 M$_\odot$ primary starts its RLOF. A quasi-conservative RLOF turns the 9 M$_\odot$ secondary into a 19 M$_\odot$ rejuvenated O-type star. When the primary remnant finally explodes as a supernova, the 19 M$_\odot$ O-type star most likely becomes a single star but the magnitude and direction of its space velocity may have changed completely, even so that its direction does not hit the parent cluster any longer. 

\subsection{Stellar dynamics in young dense star clusters}

Ultra Luminous X-ray sources (ULX) are point sources with X-ray luminosities up to 10$^{42}$ erg s$^{-1}$. MGG-11 is a young dense star cluster with Solar type metallicity $\sim$200 pc from the centre of the starburst galaxy M82, the parameters of which have been studied by McCrady et al. (2003). A ULX is associated with the cluster. When the X-rays are due to Eddington limited mass accretion onto a black hole (BH) it is straightforward to show that the mass of the BH has to be at least 1000 M$_\odot$. However how to form a star with Solar metallicity and with a mass larger than 1000 M$_\odot$? Mass segregation in a dense young cluster associated with core collapse and the formation of a runaway stellar collision process was promoted by Portegies Zwart et al. (2004). Note that the latter paper mainly addressed the dynamical evolution of a dense cluster but the evolution of the very massive stellar collision product was poorly described. 

\begin{figure*}
\centering
\includegraphics[width=8cm]{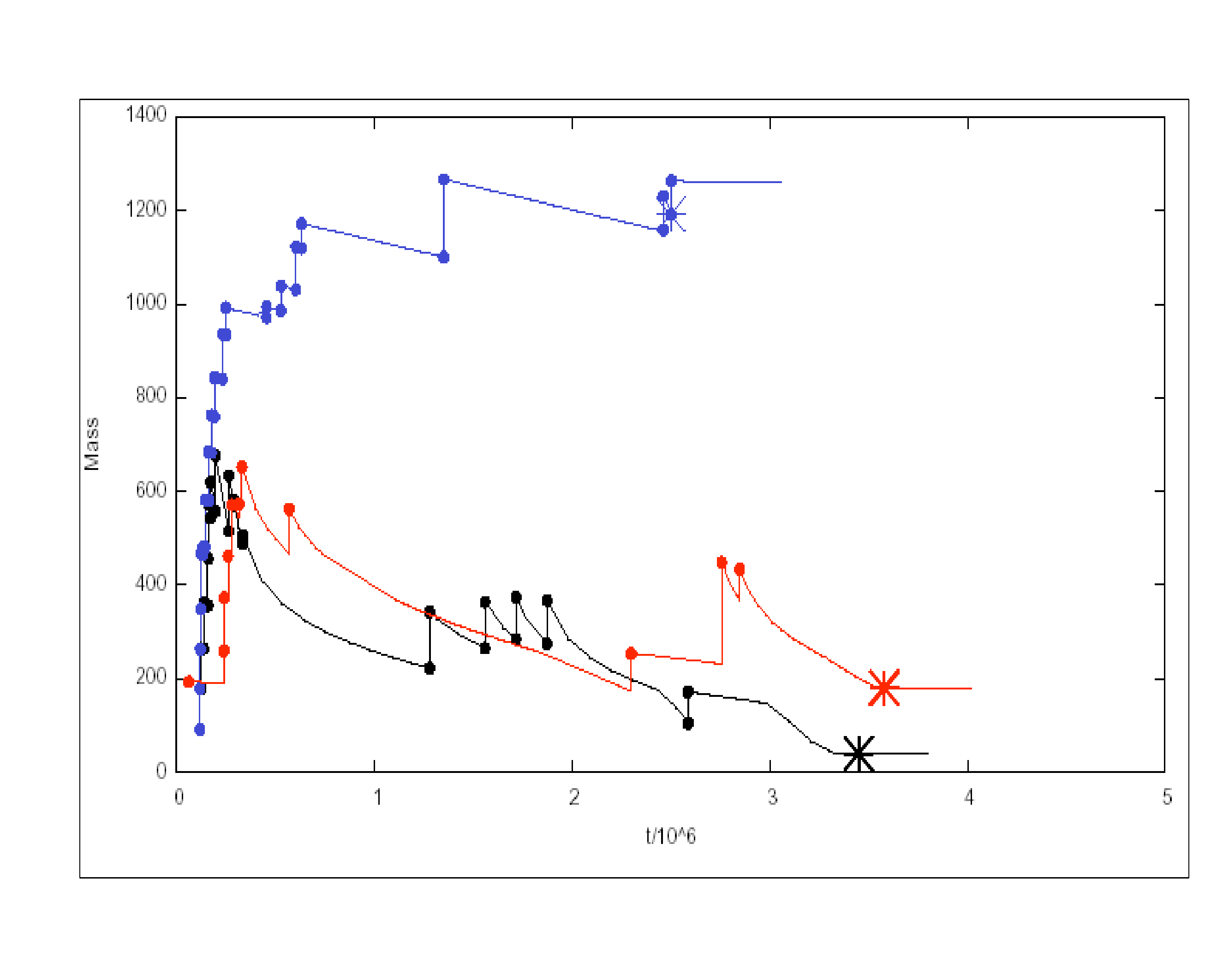}
\caption{The mass evolution of the collision runaway object in a MGG-11 type cluster. The curve with the largest final mass (Blue curve) = simulation with small mass loss, similar to the results of Portegies Zwart and McMillan (2002), the curve with the lowest final mass (Black curve) = simulation with the stellar wind mass loss formalism as discussed in Belkus et al. (2007) for Solar type metallicity, the curve with the second largest final mass (Red curve) = same as the black curve but for a SMC type metallicity (from Vanbeveren et al., 2009).}
\label{fig_dynamics}
\end{figure*}

The evolution of very massive stars has been studied in detail by Belkus et al. (2007) and Yungelson et al. (2008) and it was concluded that stellar wind mass losses during core hydrogen burning and core helium burning are very important. Belkus et al. presented a convenient evolutionary recipe for such very massive stars, which can easily be implemented in an N-body dynamical code. Our N-body code which includes this recipe has been described in Belkus (2008) and in Vanbeveren et al. (2009) and applied in order to simulate the evolution of MGG-11. In Figure 1 we show the evolution of the runaway stellar collision object for MGG-11 predicted by our code. The blue simulation is performed assuming a similar stellar wind mass loss formalism for very massive stars as the one used by Portegies Zwart et al. (2004). It can readily be checked that our simulation is very similar as the one of the latter paper and this gives some confidence that our N-body-routine is working properly. We then repeated the N-body run but with our preferred evolutionary scheme for very massive stars (the black run in Figure 1).  Our main conclusion was  the following:

\medskip
{\it \noindent Stellar wind mass loss of massive and very massive stars does not prevent the occurrence of a runaway collision event and the formation of a very massive star in a cluster like MGG-11, but after this event stellar wind mass loss during the remaining core hydrogen burning phase is large enough in order to reduce the mass again and the formation of a BH with a mass larger than $\sim$75 M$_\odot$ is rather unlikely.} 

\medskip

Similar conclusions were reached by Glebbeek et al. (2009) for MGG-11 and by Chatterjee et al. (2010) for the Arches cluster, although in both studies cluster dynamics and the evolution of very massive stars are not linked self-consistently.

Our simulations then promote the model  for the ULX in MGG-11 where the X-rays are due to super-Eddington accretion onto a stellar mass BH, a model that seems to be a most probable model for many of these  systems (Gladstone et al., 2009).

We also made a simulation for a MGG-11 like cluster but where the metallicity is significantly smaller. As can be noticed from Figure 1 (the red simulation) the formation of a BH with a mass of a few 100 M$_\odot$ is possible and this is of course a direct consequence of our adopted dependence of the stellar wind mass loss rate on the metallicity. From this simulation we are inclined to conclude that, if the progenitors of globular clusters were massive starbursts in the beginning, it is not unlikely that an intermediate mass black hole formed as a consequence of mass segregation and core collapse in a dense massive cluster.

\end{document}